\documentclass[12pt]{article}
\usepackage{enumerate}
\usepackage{amsfonts}
\usepackage{amsmath}
\usepackage{amssymb}
\usepackage{amsthm}
\usepackage{latexsym}
\usepackage{color}

\def\H{\mathcal{H}}

\def\M{\mathcal{M}}

\def\S{\mathfrak{S}}
\def\C{\mathfrak{C}}
\def\T{\mathfrak{T}}
\def\F{\mathfrak{F}}

\newcommand{\id}{\mathrm{Id}}
\newcommand{\Tr}{\mathrm{Tr}}

\newcommand{\shs}{\hspace{1pt}}
\newcounter{defin}  \newcounter{lemma}  \newcounter{theorem}
\newcounter{property} \newcounter{corol}  \newcounter{remark} \newcounter{example}
\newenvironment{lemma}{\par\refstepcounter{lemma}     \textbf{Lemma \thelemma.} }{\rm\par}
\newenvironment{theorem}{\par\refstepcounter{theorem}     \textbf{Theorem \thetheorem.}\ }{\rm\par}
\newenvironment{property}{\par\refstepcounter{property}     \textbf{Proposition \theproperty.}\ }{\rm\par}
\newenvironment{corollary}{\par\refstepcounter{corol}     \textbf{Corollary \thecorol.} }{\rm\par}
\newenvironment{definition}{\par\refstepcounter{defin}     \textbf{Definition \thedefin.}\ }{\rm\par}
\newenvironment{remark}{\par\refstepcounter{remark}     \textbf{Remark \theremark.}}{\rm\par}
\newenvironment{example}{\par\refstepcounter{example}     \textbf{Example \theexample.}}{\rm\par}

\begin{document}

\title{On quantum states with a finite-dimensional approximation property}
%\title{Universal tight continuity bounds for characteristics of energy-constrained quantum systems and channels}
\author{M.E.~Shirokov\footnote{Steklov Mathematical Institute, Moscow, Russia, email:msh@mi.ras.ru}}
\date{}
\maketitle
\vspace{-20pt}
\begin{abstract}
We consider a class (convex set) of quantum states containing all finite rank states and infinite rank states with the sufficient rate of decreasing of eigenvalues (in particular, all Gaussian states).
Quantum states from this class are characterized by the property (called the FA-property) that allows to obtain several results concerning finite-dimensional approximation of basic entropic and information characteristics of quantum systems and channels.

We obtain a simple sufficient condition of the FA-property. We  show that this property implies finiteness of the von Neumann entropy, but leave unsolved the question concerning the converse
implication.

We obtain uniform approximation results for  characteristics depending on a pair (channel, input state) and for  characteristics depending on a pair (channel, input ensemble). We  establish
the uniform continuity of the above characteristics as functions of a channel w.r.t. the strong convergence provided that the FA-property holds either for the input state or for the average state of input ensemble.
\end{abstract}

\tableofcontents

\section{Introduction}

A state of a quantum system described by a Hilbert space $\H$ is a positive trace class operator on $\H$ with unit trace having the spectral decomposition
\begin{equation}\label{rho-s-d}
\rho=\sum_{i=1}^{+\infty}\lambda_i|\varphi_i\rangle\langle\varphi_i|,
\end{equation}
where $\{\varphi_i\}$ is an orthonormal basis of eigenvectors of $\rho$ and  $\{\lambda_i\}$ is a nonincreasing sequence of nonnegative numbers -- eigenvalues of $\rho$.
If the state $\rho$ has  infinite rank then it can be approximated by the finite rank states
\begin{equation}\label{rho-r}
\rho_r=\sum_{i=1}^{r}\lambda^r_i|\varphi_i\rangle\langle\varphi_i|, \qquad \lambda^r_i=\lambda_i \left[\sum_{i=1}^{r}\lambda_i\right]^{-1}.
\end{equation}

Many characteristics used in quantum information theory have the form of a function
$$
f(\rho\shs|\shs P_1,...,P_n)
$$
on the set $\S(\H)$ of states depending on some parameters  $P_1,...,P_n$ (other states, quantum channels, quantum measurements, etc.). Naturally, the question arises under what conditions
$$
f(\rho_r\shs|\shs P_1,...,P_n)\quad \textrm{tends to}\quad f(\rho\shs|\shs P_1,...,P_n)\quad \textrm{as} \;\; r\to+\infty
$$
uniformly on $P_1,...,P_n$. It turns out that for a wide class of functions $f$ the positive answer to this question can be given in terms of the spectrum $\{\lambda_i\}$
of the state $\rho$, namely, by showing the existence of a sequence $\{g_i\}$ of nonnegative numbers such that
\begin{equation}\label{FA}
\sum_{i=1}^{+\infty}\lambda_i g_i<+\infty\quad \textrm{and} \quad \lim_{\beta\to 0^+}\left[\sum_{i=1}^{+\infty}e^{-\beta g_i}\right]^{\beta}=1.
\end{equation}

In this paper we explore the class of quantum states possessing the above property (called the FA-property in what follows). We denote this class by $\S_{\rm \textsf{FA}\!}(\H)$. By using the result from \cite{W} it is easy to show that any state in $\S_{\rm \textsf{FA}\!}(\H)$ has finite von Neumann entropy. Moreover, at this moment we can not find a state with
finite entropy for which the FA-property (\ref{FA}) is shown to be not valid. So, we may conjecture that $\S_{\rm \textsf{FA}\!}(\H)$ coincides with the set of states with finite entropy.
Fortunately, there is a simple sufficient condition for the FA-property: it holds for a state $\rho$ with the spectrum $\{\lambda_i\}$ provided that
$$
\sum_{i=1}^{+\infty}\lambda_i \ln^q i<+\infty\quad \textrm{for some} \quad q>2.
$$
This condition shows that the FA-property holds for all states whose eigenvalues tends to zero faster than $[i\ln^q i]^{-1}$ for some $q>3$, in particular, it holds
for all Gaussian states playing essential role in quantum information theory.

In Section 3 we obtain several results concerning the FA-property. We prove, in particular, that this property is preserved by a convex mixture. We also show that the FA-property
holds for a multipartite state provided that the same property holds for all its marginal states (the converse is not true, but holds for product states).

In Sections 4-5 we consider applications of the above approximation result and other attractive features of states having the FA-property.

We show that many characteristics of a quantum channel $\Phi$ having the form of a function $f(\Phi,\rho)$, where $\rho$ is an input state, possess the following uniform approximation property: if the FA-property holds for a state $\rho$ then the sequence $f(\Phi,\rho_r)$, where $\rho_r$ is the state defined in (\ref{rho-r}), tends to $f(\Phi,\rho)$ uniformly on the set of
all quantum channels  between given systems $A$ and $B$. Two versions  of the above approximation property  are obtained for characteristics having the form of a function $f(\Phi,\mu)$, where $\mu$ is an ensemble (discrete or continuous) of input states.  In this case a sufficient condition of the uniform approximation of $f(\Phi,\mu)$
is the FA-property of the average state $\bar{\rho}(\mu)$ of input ensemble $\mu$.

We also show that many characteristics of a quantum channel $\Phi$ having the form of a function $f(\Phi,\rho)$, where $\rho$ is an input state, possess the following "robustness" property:
the function $\Phi\mapsto f(\Phi,\rho)$ is uniformly continuous on the set of all quantum channels  between given systems $A$ and $B$ w.r.t. some metric generating the
strong convergence provided that
the FA-property holds for the state $\rho$. The similar property is proved for characteristics having the form of a function $f(\Phi,\mu)$, where $\mu$
is an ensemble (discrete or continuous) of input states. In this case a sufficient condition of the uniform continuity of a function $\Phi\mapsto f(\Phi,\mu)$
is the FA-property of the average state $\bar{\rho}(\mu)$ of input ensemble $\mu$.

\section{Preliminaries}

\subsection{Basic notations}

Let $\mathcal{H}$ be a separable Hilbert space,
$\mathfrak{B}(\mathcal{H})$ the algebra of all bounded operators on $\mathcal{H}$ with the operator norm $\|\cdot\|$ and $\mathfrak{T}( \mathcal{H})$ the
Banach space of all trace-class
operators on $\mathcal{H}$  with the trace norm $\|\!\cdot\!\|_1$. Let
$\mathfrak{S}(\mathcal{H})$ be  the set of quantum states (positive operators
in $\mathfrak{T}(\mathcal{H})$ with unit trace) \cite{H-SCI,H-SSQT,N&Ch,Wilde,Wilde-new}.

Denote by $I_{\mathcal{H}}$ the unit operator on a Hilbert space
$\mathcal{H}$ and by $\id_{\mathcal{\H}}$ the identity
transformation of the Banach space $\mathfrak{T}(\mathcal{H})$.\smallskip

The \emph{Bures distance} between quantum states $\rho$ and $\sigma$ is defined as
\begin{equation}\label{B-d-s}
  \beta(\rho,\sigma)=\sqrt{2\left(1-\|\sqrt{\rho}\sqrt{\sigma}\|_1\right)}.
\end{equation}
The following relations  between the Bures distance and the  trace norm hold (see \cite{H-SCI,Wilde})
\begin{equation}\label{B-d-s-r}
\textstyle\frac{1}{2}\|\rho-\sigma\|_1\leq\beta(\rho,\sigma)\leq\sqrt{\|\rho-\sigma\|_1}.
\end{equation}

The \emph{von Neumann entropy} of a quantum state
$\rho \in \mathfrak{S}(\H)$ is  defined by the formula
$H(\rho)=\operatorname{Tr}\eta(\rho)$, where  $\eta(x)=-x\log x$ for $x>0$
and $\eta(0)=0$. It is a concave lower semicontinuous function on the set~$\mathfrak{S}(\H)$ taking values in~$[0,+\infty]$ \cite{H-SCI,L-2,W}.
The von Neumann entropy satisfies the inequality
\begin{equation}\label{w-k-ineq}
H(p\rho+(1-p)\sigma)\leq pH(\rho)+(1-p)H(\sigma)+h_2(p)
\end{equation}
valid for any states  $\rho$ and $\sigma$ in $\S(\H)$ and $p\in(0,1)$, where $\,h_2(p)=\eta(p)+\eta(1-p)\,$ is the binary entropy \cite{N&Ch,Wilde}.\smallskip

The \emph{quantum relative entropy} for two states $\rho$ and
$\sigma$ in $\mathfrak{S}(\mathcal{H})$ is defined as
$$
H(\rho\shs\|\shs\sigma)=\sum\langle
\varphi_i|\,\rho\log\rho-\rho\log\sigma\,|\varphi_i\rangle,
$$
where $\{|\varphi_i\rangle\}$ is the orthonormal basis of
eigenvectors of the state $\rho$ and it is assumed that
$H(\rho\shs\|\shs\sigma)=+\infty$ if $\,\mathrm{supp}\rho\shs$ is not
contained in $\shs\mathrm{supp}\shs\sigma$ \cite{H-SCI,L-2}.\footnote{The support $\mathrm{supp}\rho$ of a state $\rho$ is the closed subspace spanned by the eigenvectors of $\rho$ corresponding to its positive eigenvalues.}\smallskip

A finite or
countable collection $\{\rho_{k}\}$ of quantum states
with a  probability distribution $\{p_{k}\}$ is called (discrete) \emph{ensemble} and denoted by $\{p_k,\rho_k\}$. The state $\bar{\rho}=\sum_{k} p_k\rho_k$ is called  the \emph{average state} of  $\{p_k,\rho_k\}$.  The Holevo quantity of an ensemble
$\{p_k,\rho_k\}$ is defined as
\begin{equation}\label{H-q}
\chi(\{p_k,\rho_k\})= \sum_{k} p_k H(\rho_k\|\bar{\rho})=H(\bar{\rho})-\sum_{k} p_kH(\rho_k),
\end{equation}
where the second formula is valid if $H(\bar{\rho})$ is finite. This quantity is a upper bound on the classical information obtained from  quantum measurements
over the ensemble \cite{H-73}.\smallskip

Let $G$  be a  positive operator on the space $\H$. For any positive operator $\rho$ in $\T(\H)$
we assume that the value of $\Tr G\rho$ (finite or infinite) is defined as $\sup_n\Tr P_nG\rho$, where $P_n$ is the spectral projector of $G$ corresponding to the interval $[0,n]$.\smallskip

If the operator $G$ satisfies the condition
\begin{equation}\label{H-cond}
  \mathrm{Tr}\, e^{-\beta G}<+\infty,\quad \forall \beta>0
\end{equation}
then the entropy is continuous and bounded on the set $\C_{G,E}$ of states $\rho$ in $\S(\H)$ determined by the
inequality $\Tr G\rho\leq E$ for any $E>g_{\rm m}$, where $g_{\rm m}$ is the infimum of the spectrum of $G$ \cite{W}. By Proposition 1 in \cite{EC} condition (\ref{H-cond}) is equivalent to
the following one
\begin{equation}\label{F-fun}
F_{G}(E)\doteq\sup_{\rho\in\C_{G,E}}H(\rho)=o(E)\quad \textrm{as} \;\; E\to+\infty.
\end{equation}

Condition (\ref{H-cond}) implies that the operator $G$ has discrete spectrum of finite multiplicity.
In Dirac's notations any such operator ${G}$ can be represented as follows
\begin{equation}\label{H-rep}
G=\sum_{i=1}^{+\infty} g_i|\tau_i\rangle\langle \tau_i|
\end{equation}
on the domain $\mathcal{D}(G)=\{ \varphi\in\H\,|\,\sum_{i=1}^{+\infty} g^2_i|\langle\tau_i|\varphi\rangle|^2<+\infty\}$, where
$\left\{\tau_i\right\}_{i=1}^{+\infty}$ is the orthonormal basis of eigenvectors of $G$
corresponding to the nondecreasing sequence $\left\{\smash{g_i}\right\}_{i=1}^{+\infty}$ of eigenvalues
tending to $+\infty$.\smallskip

The restriction (\ref{F-fun}) on the rate of increasing of the function $F_{G}(E)$ can be strengthened by assuming that
\begin{equation}\label{H-cond+}
  \lim_{\beta\rightarrow0^+}\left[\mathrm{Tr}\, e^{-\beta G}\right]^{\beta}=1.
\end{equation}
By Lemma 1 in \cite{AFM} this condition is equivalent to
the following one
\begin{equation}\label{F-fun+}
F_{G}(E)=o(\sqrt{E})\quad \textrm{as} \;\; E\to+\infty.
\end{equation}
Condition (\ref{H-cond+}) is slightly stronger than condition (\ref{H-cond}). In terms of the sequence $\{g_i\}$ of eigenvalues of $G$
condition (\ref{H-cond}) means that $\,\lim_{i\rightarrow\infty}g_i/\ln i=+\infty$, while condition (\ref{H-cond+}) is valid  if $\;\liminf_{i\rightarrow\infty} g_i/\ln^q i>0\,$ for some $\,q>2$ \cite[Proposition 1]{AFM}.\smallskip

A \emph{quantum operation} $\,\Phi$ from a system $A$ to a system
$B$ is a completely positive trace-non-increasing linear map from
$\mathfrak{T}(\mathcal{H}_A)$ into $\mathfrak{T}(\mathcal{H}_B)$, a trace preserving quantum operation is called \emph{quantum channel} \cite{H-SCI,Wilde}.
For any  quantum operation $\,\Phi:A\rightarrow B\,$ the Stinespring theorem implies the existence of a Hilbert space
$\mathcal{H}_E$ and  a contraction
$V_{\Phi}:\mathcal{H}_A\rightarrow\H_{BE}\doteq\mathcal{H}_B\otimes\mathcal{H}_E$ such
that
\begin{equation*}%\label{St-rep}
\Phi(\rho)=\mathrm{Tr}_{E}V_{\Phi}\rho V_{\Phi}^{*},\quad
\rho\in\mathfrak{T}(\mathcal{H}_A).
\end{equation*}
The operation
\begin{equation}\label{c-ch}
\widehat{\Phi}(\rho)=\mathrm{Tr}_{B}V_{\Phi}\rho V_{\Phi}^{*},\quad
\rho\in\mathfrak{T}(\mathcal{H}_A),
\end{equation}
from the system $A$ to the system $E$ associated with the space $\H_E$ is called \emph{complementary} to the operation $\Phi$. The complementary operation is uniquely defined up to the \emph{isometrical equivalence}, i.e. if $\,\widehat{\Phi}':A\rightarrow E'$ is the operation defined by formula  (\ref{c-ch})
via some other contraction  $\,V'_{\Phi}:\H_A\rightarrow\H_{BE'}\,$ then there exists a partial isometry
$W:\H_E\rightarrow\H_{E'}$ such that $\widehat{\Phi}'(\rho)=W\widehat{\Phi}(\rho)W^*$ and $\widehat{\Phi}(\rho)=W^*\widehat{\Phi}'(\rho)W$ for all $\rho\in \T(\H_A)$ \cite{H-SCI, H-c-ch}.
If $\Phi$ is a channel then $V_{\Phi}$ is an isometry and hence $\widehat{\Phi}$ is a channel as well.
\smallskip

A quantum operation $\,\Phi$ from a system $A$ to a system
$B$ preserves finiteness of the entropy if
$$
H(\Phi(\rho))\doteq [\Tr\Phi(\rho)]H\!\left(\frac{\Phi(\rho)}{\Tr\Phi(\rho)}\right)<+\infty
$$
for any state $\rho$ in $\S(\H_A)$ such that $H(\rho)<+\infty$. By Theorem 1 in \cite{PFE} this
property implies continuity of the function $\rho\mapsto H(\Phi(\rho))$ on any subset of $\S(\H_A)$ where the entropy is continuous, it is equivalent to the finiteness
of
\begin{equation}\label{HM}
  H_{\rm max}^{\rm p}(\Phi)=\sup_{\rho\in\mathrm{ext}\S(\H_A)}H(\Phi(\rho)),
\end{equation}
where the supremum is over the set $\mathrm{ext}\S(\H_A)$ of all pure input states. The class of channels  preserving finiteness of the entropy contains channels with finite Choi rank and some other
types of channels (see a detailed classification in \cite[Section 3.3]{PFE}). \smallskip

Let $G$ be a  positive operator on the Hilbert space $\H_A$ (treated as a Hamiltonian of the input system $A$). The \emph{energy-constrained Bures distance}
\begin{equation}\label{ec-b-dist}
\beta_G^E(\Phi,\Psi)=\sup_{\rho\in\S(\H_{AR}), \Tr G\rho_A\leq E} \beta(\Phi\otimes \id_R(\rho),\Psi\otimes \id_R(\rho)), \quad E> g_{\rm m},
\end{equation}
between quantum channels $\Phi$  and $\Psi$ from $A$ to $B$, where $\beta$ in the r.h.s. is the Bures distance between quantum states defined in (\ref{B-d-s}), $\,R\,$ is an infinite-dimensional quantum system and $g_{\rm m}$ is the infimum of the spectrum of $G$, is a metric on the set
of all quantum channels from $A$ to $B$. Its properties are described in Proposition 1  in \cite{CID}. Calculations of $\beta_G^E(\Phi,\Psi)$
for real quantum channels can be found in \cite{Nair}.

\section{The FA-property and its use}

In what follows we denote by $\{\lambda^{\rho}_i\}$ the non-increasing sequence of eigenvalues of a state $\rho$ in $\S(\H)$ taking  the multiplicity into account.\smallskip

\begin{definition}\label{FA-def} The \emph{FA-property} holds for a state $\rho$ if there exists a sequence $\{g_i\}$ of nonnegative numbers  such that
\begin{equation}\label{FA+}
\sum_{i=1}^{+\infty}\lambda^{\rho}_i g_i<+\infty\quad \textrm{and} \quad \lim_{\beta\to 0^+}\left[\sum_{i=1}^{+\infty}e^{-\beta g_i}\right]^{\beta}=1.
\end{equation}
\end{definition}

We will denote the set of all states in $\S(\H)$ having the FA-property by $\S_{\rm \textsf{FA}\!}(\H)$.
\smallskip

It is easy to show that the FA-property holds for all finite rank states and for all states whose spectrum forms a geometric series, in particular, for all Gaussian
states (in the last case one can take the sequence $g_i=i$). The following theorem states, in particular, that a quantum state has the FA-property if and only if it has a finite energy w.r.t.
some "Hamiltonian" with sufficiently high increasing rate of the spectrum.\smallskip

\begin{theorem}\label{FA-ch}
\emph{Let $\rho$ be a state in $\S(\H)$. The following
properties are equivalent:}
\begin{enumerate}[(i)]
  \item \emph{the FA-property holds for the state $\rho$; }
  \item \emph{there is a positive operator $G$ on $\H$ satisfying condition (\ref{H-cond+}) s.t. $\Tr G\rho<+\infty$;}
  \item \emph{$H(\rho)<+\infty\,$ and $\,H(\rho\shs\|\shs\sigma)<+\infty\,$  for some state $\sigma$ in $\S(\H)$ such that}
\begin{equation}\label{sigma-c}
\Tr \sigma^q<+\infty\;\;\forall q>0\quad  \textit{and} \quad \lim_{q\to 0^+} \left[\Tr \sigma^q\right]^q=1.
\end{equation}
\end{enumerate}

\emph{The equivalent properties  $(i)\textrm{-}(iii)$ hold provided that}
\begin{equation}\label{sp-cond}
\sum_{i=1}^{+\infty}\lambda^{\rho}_i \ln^q i<+\infty\quad \textit{for some} \quad q>2.
\end{equation}
\end{theorem}\smallskip

\begin{remark}\label{r1+}
Condition (\ref{sigma-c}) holds for states with rapidly decreasing spectrum, for example,
for states whose spectrum forms a geometric series.\smallskip
\end{remark}

\emph{Proof.} Let $\rho=\sum_{i=1}^{+\infty}\lambda^{\rho}_i|\varphi_i\rangle\langle \varphi_i|$ be the spectral decomposition of the state $\rho$.\smallskip

$\mathrm{(i)}\Rightarrow\mathrm{(ii)}$.  Let $\{g_i\}$ be a sequence of nonnegative numbers such that (\ref{FA+}) holds.
Then the positive operator
\begin{equation}\label{DH}
G=\sum_{i=1}^{+\infty}g_i|\varphi_i\rangle\langle \varphi_i|
\end{equation}
on $\H$ satisfies condition (\ref{H-cond+}) and $\Tr G\rho=\sum_{i=1}^{+\infty}\lambda^{\rho}_i g_i<+\infty$.\smallskip

$\mathrm{(ii)}\Rightarrow\mathrm{(i)}$. Since the operator $G$ satisfies condition (\ref{H-cond+}), it has discrete spectrum of finite multiplicity. So, we may assume that it has the form (\ref{H-rep}), in which
$\{g_i\}$ is the nondecreasing sequence of eigenvalues of $G$ such that $\lim_{\beta\to 0^+}\left[\sum_{i=1}^{+\infty}e^{-\beta g_i}\right]^{\beta}=1$. Then
$$
\Tr G\rho=\sum_{i=1}^{+\infty} g_i\mu_i<+\infty, \quad \textrm{where}\quad \mu_i=\langle \tau_i|\rho|\tau_i\rangle\;\;\forall i.
$$
By the Schur–Horn theorem (cf.\cite{S-T-1,S-T-2}) the sequence $\{\mu_i\}$ is majorised by the sequence $\{\lambda^{\rho}_i\}$. It follows that
\begin{equation}\label{mr-cor}
  S_n^\lambda\doteq\sum_{i\geq n}\lambda^{\rho}_i\leq S_n^\mu\doteq\sum_{i\geq n}\mu^{\downarrow}_i\quad  n=1,2,...,
\end{equation}
where $\{\mu^{\downarrow}_i\}$ is the sequence obtained by rearranging the sequence $\{\mu_i\}$ in the non-increasing order.
Since
$$
\sum_{i=1}^{+\infty}\lambda^{\rho}_i g_i=\sum_{n=1}^{+\infty}S_n^\lambda d_n\quad \textrm{and}  \quad \sum_{i=1}^{+\infty}\mu^{\downarrow}_i g_i=\sum_{n=1}^{+\infty}S_n^\mu d_n,
$$
where $d_1=g_1$ and $d_n=g_{n}-g_{n-1}$ for all $n>1$, it follows from (\ref{mr-cor}) and the
easily verified inequality
$$
\sum_{i=1}^{+\infty} g_i\mu^{\downarrow}_i\leq \sum_{i=1}^{+\infty} g_i\mu_i
$$
that the finiteness of the series $\sum_{i=1}^{+\infty} g_i\mu_i$ implies the finiteness of the series $\sum_{i=1}^{+\infty} g_i\lambda^{\rho}_i$.\smallskip

$\mathrm{(i)}\Rightarrow\mathrm{(iii)}$. Let $\{g_i\}$ be a sequence of nonnegative numbers such that (\ref{FA+}) holds.
Since the state $\rho$ has finite energy w.r.t. the "Hamiltonian" (\ref{DH}) satisfying condition (\ref{H-cond}) it has a finite entropy \cite{W}.
Consider the state
$$
\sigma=c^{-1}\sum_{i=1}^{+\infty} e^{-g_i}|\varphi_i\rangle\langle \varphi_i|,\quad c=\sum_{i=1}^{+\infty} e^{-g_i},
$$
on $\S(\H)$ satisfying condition (\ref{sigma-c}). Then $-\Tr \rho\ln\sigma=\ln c+\sum_{i=1}^{+\infty}g_i\lambda^{\rho}_i<+\infty$.
Since $H(\rho)<+\infty$ this implies that $H(\rho\shs\|\shs\sigma)<+\infty$.\smallskip

$\mathrm{(iii)}\Rightarrow\mathrm{(ii)}$. The finiteness of $H(\rho)$ and $H(\rho\shs\|\shs\sigma)$
implies the finiteness of $\Tr G_{\sigma}\rho$, where  $G_{\sigma}=-\ln\sigma$. By the conditions in (\ref{sigma-c})
the operator $ G_{\sigma}$ satisfies condition (\ref{H-cond+}).

\smallskip

The last assertion of the theorem  follows from  Lemma 2 in \cite{AFM} stating that the sequence $\,g_i=\ln^q i\,$ satisfies the second condition in (\ref{FA+}) if and only if $\,q>2$.
$\square$
\smallskip

By Theorem \ref{FA-ch} any state with the FA-property
has finite entropy. But the converse statement is an interesting \textbf{open question}: at this moment we can neither prove that the finiteness of the von Neumann entropy implies the
FA-property nor prove that the FA-property is not valid for at least one state with finite entropy.\footnote{I would be grateful for any comments concerning this question.}

By Proposition 4 in \cite{EC} the finiteness of the von Neumann entropy of a state $\rho$  \emph{is equivalent} to the existence of a sequence $\{g_i\}$ of nonnegative numbers  such that
\begin{equation*}%\label{FA+}
\sum_{i=1}^{+\infty}\lambda^{\rho}_i g_i<+\infty\quad \textrm{and} \quad \sum_{i=1}^{+\infty}e^{-\beta g_i}<+\infty\quad \forall \beta>0.
\end{equation*}
The arguments from the proof of Theorem \ref{FA-ch} shows that the  existence of such a sequence for a given state $\rho$ is equivalent to the  existence of a positive operator
$G$ satisfying condition (\ref{H-cond}) such that $\Tr G\rho<+\infty$.\smallskip

By Theorem  \ref{FA-ch} a state $\rho$ has the FA-property if its spectrum $\{\lambda^{\rho}_i\}$ has the sufficient rate of decreasing, namely $\lambda^{\rho}_i=o([i\ln^{q}i]^{-1})$ for some $q>3$.
A state $\rho$ with the spectrum $\lambda^{\rho}_i\sim[i\ln^{3}i]^{-1}$ has finite entropy but does not satisfy  condition (\ref{sp-cond}). Lemma 2 in \cite{AFM} gives a reason
to assume that the FA-property does not hold for any such state, but we have not found a rigorous proof of this conjecture.
\smallskip

According to the terminology accepted in \cite{Har} a quantum state $\rho$ is called \emph{power dominated} if $\Tr \rho^{\shs q}<+\infty$ for some $q<1$
and \emph{hyperbolic} if $\Tr \rho^{\shs q}=+\infty$ for all $q<1$. It is easy to see that any power dominated state has finite entropy (but the converse is not true). Theorem \ref{FA-ch}
implies the following \smallskip

\begin{corollary}\label{FA-sc-c}
\emph{The FA-property holds for any power dominated state $\rho$ in $\S(\H)$.}
\end{corollary}\smallskip

\emph{Proof.} Assume that $\Tr \rho^{\shs q}=\sum_{i=1}^{+\infty}[\lambda^{\rho}_i]^q<+\infty$, where $q\in(0,1)$. Since
$\sum_{i=1}^{+\infty}\lambda^{\rho}_i=1$, we have $\lambda^{\rho}_i\leq 1/i$ for each $i$. It follows that  $[\lambda^{\rho}_i]^{1-q} \ln^3 i$ tends to zero as $i\to+\infty$.
Hence
$$
\sum_{i=1}^{+\infty}\lambda^{\rho}_i \ln^3 i= \sum_{i=1}^{+\infty}[\lambda^{\rho}_i]^q \left[[\lambda^{\rho}_i]^{1-q}\ln^3 i\right]<+\infty.\; \square
$$

A quantum state $\rho$ majorises  a quantum state $\sigma$  if
\begin{equation*}%\label{mr-cor}
  \sum_{i=1}^n\lambda^{\rho}_i\geq \sum_{i=1}^n\lambda_i^{\sigma}\quad  n=1,2,...
\end{equation*}
(according to our notation the sequences $\{\lambda^{\rho}_i\}$ and $\{\lambda^{\sigma}_i\}$ of eigenvalues are arranged in the non-increasing order) \cite{S-T-1}.
If this holds we will write $\rho\succ\sigma$. \smallskip

\begin{corollary}\label{FA-sc} \emph{The FA-property  holds for a state $\rho$ in $\S(\H)$ if there is a state
$\sigma$ in $\S(\H)$ with  the FA-property such that either $\rho\succ\sigma$ or $\rho\leq c\sigma$ for some $c>0$.}
\end{corollary} \smallskip

\emph{Proof.} Assume that $\rho\succ\sigma$. Let $\{g_i\}$ be a nondecreasing sequence of nonnegative numbers such that (\ref{FA+}) holds
with $\{\lambda^{\rho}_i\}$ replaced by $\{\lambda^{\sigma}_i\}$. By using  the arguments from the proof of the implication $\mathrm{(ii)}\Rightarrow\mathrm{(i)}$ in Theorem \ref{FA-ch}
it is easy to show that
$$
\sum_{i=1}^{+\infty}g_i\lambda^{\rho}_i\leq \sum_{i=1}^{+\infty}g_i\lambda^{\sigma}_i<+\infty.
$$

If $\rho\leq c\sigma$ for some $c>0$ then the validity of the FA-property for the state $\rho$
can be easily shown by using the equivalence of properties $\rm (i)$ and $\rm (ii)$ in Theorem \ref{FA-ch}. $\square$ \smallskip

\begin{property}\label{new-FAP} A) \emph{The set $\,\S_{\rm \textsf{FA}\!}(\H)$ of  states having the FA-property is convex.}\smallskip

%B) \emph{If $\rho_{12}$ is a state in $\S(\H_1\otimes\H_2)$ such that  $\rho_{1}\in\S_{\rm \textsf{FA}\!}(\H_1)$ and $\rho_{2}\in\S_{\rm \textsf{FA}\!}(\H_2)$ then
%$\rho_{12}\in\S_{\rm \textsf{FA}\!}(\H_1\otimes\H_2)$}.

B) \emph{If $\rho_{1...n}$ is a state in $\S(\H_1\otimes...\otimes\H_n)$ such that  $\rho_{k}\in\S_{\rm \textsf{FA}\!}(\H_k)$ for all $\,k=1,..,n$ then
$\,\rho_{1...n}\in\S_{\rm \textsf{FA}\!}(\H_1\otimes...\otimes\H_n)$.\footnote{$\rho_{k}$ is the marginal state of $\rho_{1..n}$ corresponding to the space $\H_k$.}  A state $\rho_1\otimes...\otimes\rho_n$ belongs to the set $\,\S_{\rm \textsf{FA}\!}(\H_1\otimes...\otimes\H_n)$
if and only if $\,\rho_{k}\in\S_{\rm \textsf{FA}\!}(\H_k)$ for all $\,k=1,..,n$.}
\end{property} \smallskip

\emph{Proof.} A) Let  $\rho$ and $\sigma$ be arbitrary states in $\S_{\rm \textsf{FA}\!}(\H)$. Let $\{g^{\rho}_i\}$ and $\{g^{\sigma}_i\}$ be sequences of nonnegative numbers such that (\ref{FA+}) holds for the spectrums of $\rho$ and $\sigma$.

Let $\omega=\frac{1}{2}(\rho+\sigma)$. By Weyl's inequality we have\footnote{The validity of Weyl's inequality for trace class operators follows, f.i., from Theorem 5.10 in \cite{Markus}.}
\begin{equation}\label{W-ineq}
2\lambda^{\omega}_{2i-1}\leq \lambda^{\rho}_i+\lambda^{\sigma}_i,\quad 2\lambda^{\omega}_{2i}\leq \lambda^{\rho}_{i+1}+\lambda^{\sigma}_i\leq \lambda^{\rho}_i+\lambda^{\sigma}_i,\quad i=1,2,...
\end{equation}
Let $\{\tilde{g}_i\}_{i=1}^{+\infty}$ be the sequence defined by setting
$\tilde{g}_{2i-1}=\tilde{g}_{2i}=g^{\rho}_i$ if $\lambda^{\rho}_i\geq\lambda^{\sigma}_i$ and $\tilde{g}_{2i-1}=\tilde{g}_{2i}=g^{\sigma}_i$ otherwise.
Then it follows from (\ref{W-ineq}) that
$$
\sum_{i=1}^{+\infty}\tilde{g}_i\lambda^{\omega}_i=\!\sum_{i=1}^{+\infty}(\tilde{g}_{2i-1}\lambda^{\omega}_{2i-1}+\tilde{g}_{2i}\lambda^{\omega}_{2i})
\leq\!\sum_{i=1}^{+\infty}(\lambda^{\rho}_i+\lambda^{\sigma}_i)\frac{(\tilde{g}_{2i-1}+\tilde{g}_{2i})}{2}\leq2\!\sum_{i=1}^{+\infty}(\lambda^{\rho}_ig^{\rho}_i+\lambda^{\sigma}_ig^{\sigma}_i)<+\infty.
$$
Since
$$
\sum_{i=1}^{+\infty}e^{-\beta\tilde{g}_i}=\sum_{i=1}^{+\infty}(e^{-\beta\tilde{g}_{2i-1}}+e^{-\beta\tilde{g}_{2i}})\leq 2\sum_{i=1}^{+\infty}(e^{-\beta g^{\rho}_{i}}+e^{-\beta g^{\sigma}_{i}}),
$$
by using the inequality $(x+y)^{\beta}\leq \max\{[2x]^{\beta},[2y]^{\beta}\}$ it is easy to show that
$$
\lim_{\beta\to 0^+}\left[\sum_{i=1}^{+\infty}e^{-\beta \tilde{g}_i}\right]^{\beta}=1.
$$

B) It is sufficient to prove both assertions in the case $n=2$.

By Theorem \ref{FA-ch} there exist positive operators $G_1$ on $\H_1$ and $G_2$ on $\H_2$ satisfying condition (\ref{H-cond+}) such that $\Tr G_1\rho_1$ and $\Tr G_2\rho_2$ are finite. To show that the state $\rho_{12}$ has the FA-property it suffices, by Theorem \ref{FA-ch}, to prove that the operator $G_{12}=G_1\otimes I_{\H_2}+I_{\H_1}\otimes G_2$
on $\H_1\otimes\H_2$ satisfies condition (\ref{H-cond+}). This can be done by using the equivalence of (\ref{H-cond+}) and (\ref{F-fun+}), since the subadditivity of the entropy implies that
$$
F_{G_{12}}(E)\leq\max_{E_1+E_2\leq E\,}[F_{G_{1}}(E_1)+F_{G_{2}}(E_2)]\leq F_{G_{1}}(E)+F_{G_{2}}(E).
$$

To prove the second assertion it suffices to show that the FA-property of a state $\rho_1\otimes\rho_2$ implies the same
property of the states $\rho_1$ and $\rho_2$.  The spectrum of the state $\rho_1\otimes\rho_2$ can be represented as
the double sequence $\{\lambda^{\rho_1}_i\lambda^{\rho_2}_j\}_{i,j\geq 1}$. So, if the FA-property holds for this state then there exists  a double  sequence $\{g_{ij}\}_{i,j\geq 1}$ of nonnegative numbers such that
\begin{equation}\label{FA+sp}
\sum_{i,j=1}^{+\infty}\lambda^{\rho_1}_i\lambda^{\rho_2}_j g_{ij}<+\infty\quad \textup{and} \quad \lim_{\beta\to 0^+}\left[\sum_{i,j=1}^{+\infty}e^{-\beta g_{ij}}\right]^{\beta}=1.
\end{equation}
Let $g^1_i=\sum_{j=1}^{+\infty}\lambda^{\rho_2}_j g_{ij}$, $i=1,2,..$. Then $\,\sum_{i=1}^{+\infty}\lambda^{\rho_1}_i g^1_{i}<+\infty\,$ by the first relation in (\ref{FA+sp}) and the convexity of the function $e^{-\beta x}$ implies
$$
\sum_{i=1}^{+\infty}e^{-\beta g^1_{i}}\leq \sum_{i=1}^{+\infty}\sum_{j=1}^{+\infty} \lambda^{\rho_2}_j e^{-\beta g_{ij}}\leq \sum_{i=1}^{+\infty}\sum_{j=1}^{+\infty} e^{-\beta g_{ij}}\quad \forall\beta>0.
$$
Using this inequality and the second relation in (\ref{FA+sp}) it is easy to show that the second relation in (\ref{FA+}) holds for the sequence $\{g^1_i\}$. Similarly, one can prove that the state $\rho_2$ has the FA-property. $\square$\smallskip

In modern convex analysis a subset $\mathcal{B}$ of a convex set $\mathcal{A}$ is called a \emph{face} if the subset $\mathcal{B}$ is convex and contains any line
segment from the set $\mathcal{A}$ that has at least one internal point belonging to $\mathcal{B}$ \cite{Roc}.\smallskip

\begin{corollary}\label{face} \emph{The set $\S_{\rm \textsf{FA}\!}(\H)$ is a dense face of $\S(\H)$.}
\end{corollary} \smallskip

\emph{Proof.} The set $\S_{\rm \textsf{FA}\!}(\H)$ is dense in $\S(\H)$, since it contains all finite-rank states.
The convexity of $\S_{\rm \textsf{FA}\!}(\H)$ follows from Proposition \ref{new-FAP}.

Assume that $\rho$ is a state in $\S_{\rm \textsf{FA}\!}(\H)$ and $\rho=(1-p)\rho_1+p\rho_2$, where $\rho_i\in\S(\H)$, $i=1,2$, and $p\in(0,1)$.
Then $(1-p)\rho_1\leq\rho$ and $p\rho_2\leq\rho$. Hence  $\rho_i\in\S_{\rm \textsf{FA}\!}(\H)$, $i=1,2$ by Corollary \ref{FA-sc}. $\square$\smallskip

\begin{remark}\label{r1} It is interesting to note that all the assertions of Proposition \ref{new-FAP} and Corollaries \ref{FA-sc-c},\ref{FA-sc} and \ref{face} are valid with the FA-property replaced by
"the entropy finiteness" property of a state (respectively, with the set $\S_{\rm \textsf{FA}\!}(\H)$ replaced by the set of states in $\S(\H)$ with finite entropy, etc).  But in the FA-property case the proofs of these assertions require more technical efforts.
\end{remark}\medskip

Now we turn to the main technical result related to the FA-property.
\smallskip

Let $L(C,T,D)$ be the class of all functions $f$ on the set
$$
\S_0(\H)\doteq\left\{\rho\in\S(\H)\,|\,H(\rho)<+\infty\shs\right\}
$$
such that
\begin{equation}\label{F-p-1}
-a_f h_2(p)\leq f(p\rho+(1-p)\sigma)-p f(\rho)-(1-p)f(\sigma)\leq b_f h_2(p)
\end{equation}
for any states $\rho$  and $\sigma$ in $\S_0(\H)$ and any $p\in(0,1)$, and
\begin{equation}\label{F-p-2}
-c^-_f H(\rho)-t^-_f\leq f(\rho)\leq c^+_f H(\rho)+t^+_f
\end{equation}
for any state $\rho$ in $\S_0(\H)$, where $h_2$ is the binary entropy (defined after (\ref{w-k-ineq}))
and $a_f$,$b_f$,$c^{\pm}_f$ and $t^{\pm}_f$ are nonegative numbers such that $a_f+b_f=D$, $c^-_f+c^+_f=C$ and $t^-_f+t^+_f=T$.\smallskip

There are several important characteristics used in quantum information theory belonging to one of the classes $L(C,T,D)$. For example,
\begin{itemize}
  \item the von Neumann entropy $H(\rho)$ belongs to the class $L(1,0,1)$. This follows from its nonnegativity, concavity and inequality (\ref{w-k-ineq}).
  \item the output entropy $H_{\Phi}(\rho)\doteq H(\Phi(\rho))$ and the entropy exchange $H(\Phi,\rho)\doteq H(\widehat{\Phi}(\rho))$
of a quantum operation $\Phi$ preserving finiteness of the entropy belong to the class $L(1,H^{\rm p}_{\rm max}(\Phi),1)$, where $H^{\rm p}_{\rm max}(\Phi)$ is defined in (\ref{HM}) (see the below Example \ref{e-1}).
 \item the mutual information $I(\Phi,\rho)$ of any quantum channel $\Phi$ belongs to the class $L(2,0,2)$ (see \cite[Section 4.3]{MCB}).
 \item the coherent information $I_c(\Phi,\rho)$ of any quantum channel $\Phi$ belongs to the class $L(2,0,2)$ (see \cite[Section 4.3]{MCB}).
 \item the information gain $IG(\mathfrak{M},\rho)$ of any quantum measurement $\mathfrak{M}$ with countable outcome set belongs to the class $L(2,0,2)$ (see the below Example \ref{e-2}).
\end{itemize}

The following theorem shows that for any function $f$ belonging to  one of the classes $L(C,T,D)$ its value $f(\rho)$ at any state $\rho$ having the FA-property can be approximated by the sequence
$\{f(\rho_r)\}_r$, where $\rho_r$ is the state of rank $r$ obtained by truncation of the spectral decomposition of $\rho$. Moreover, the theorem states that this approximation
is uniform on a given class $L(C,T,D)$. \smallskip

\begin{theorem}\label{main} \emph{Let $\rho$ be a state in $\S_{\!\rm \textsf{FA}\!}(\H)$ and $\rho_r=P_r\rho/[\Tr P_r\rho]$, where $P_r$ is the spectral projector of $\rho$ corresponding to its
$r$ maximal eigenvalues. Then there
is a natural number $r_0$ and a vanishing sequence $\{Y_{C,T\!,D}(r)\}_{r\in\mathbb{N}}$ (depending only on the spectrum of $\rho$ and the nonnegative numbers $C,T$ and $D$) such that
$$
|f(\rho_r)-f(\rho)|\leq Y_{C,T\!,D}(r)  \qquad \forall r\geq r_0
$$
for any function $f$ in $L(C,T,D)$.}
\end{theorem}\smallskip

\emph{Proof.} Let $\rho=\sum_{i=1}^{+\infty}\lambda^{\rho}_i|\varphi_i\rangle\langle \varphi_i|$ be the spectral decomposition of the state $\rho$ and $\{g_i\}$ a non-decreasing sequence of nonnegative numbers such that (\ref{FA+}) holds. Then $P_r=\sum_{i=1}^{r}|\varphi_i\rangle\langle \varphi_i|$. Let $G$ be the positive operator on $\H$ defined in (\ref{DH}) satisfying condition (\ref{H-cond+}) such that
\begin{equation}\label{E-r}
E_{\rho}\doteq \Tr G\rho=\sum_{i=1}^{+\infty}\lambda^{\rho}_i g_i<+\infty.
\end{equation}

For any function $f$ in $L(C,T,D)$ and a state $\sigma$ in $\S(\H)$ with finite $E_{\sigma}=\Tr G\sigma$ we have
\begin{equation}\label{F-p-2+}
-c^-_f F_{G}(E_{\sigma})-t^-_f\leq f(\sigma)\leq c^+_f F_{G}(E_{\sigma})+t^+_f,
\end{equation}
where $F_{G}$ is the function defined in (\ref{F-fun}).

It is easy to see  that
$$
\|\rho-\rho_r\|_1\leq 2\Tr\bar{P}_r\rho,\quad \bar{P}_r=I_{\H}-P_r.
$$

Let $r_0$ be the minimal natural number such that $g_{r_0}>E_{\rho}$. Then $\frac{1}{2}\|\rho-\rho_r\|_1\leq\delta_r\doteq \Tr\bar{P}_r\rho\leq 1$
and $\Tr G\rho_r\leq E_{\rho}$ for all $r\geq r_0$. Thus, by using (\ref{E-r}) and (\ref{F-p-2+}) we obtain from Theorem 1 in \cite{AFM} that
\begin{equation}\label{d-one}
  |f(\rho)-f(\rho_r)|\leq \sqrt{2\delta_r}\left[CF_{G}\!\left(E_{\rho}/\delta_r\right)+T\right]+Dg(\sqrt{2\delta_r}) \qquad \forall r\geq r_0
\end{equation}
for any function $f\in L(C,T,D)$, where $g(x)=(1+x)h_2(x/(1+x))$.  \smallskip

Since $\delta_r$ tends to zero as $r\to+\infty$, denoting the r.h.s. of (\ref{d-one}) by $Y_{C,T\!,D}(r)$ and using the equivalence of (\ref{H-cond+}) and (\ref{F-fun+}) we obtain the assertion of the theorem. $\square$\smallskip

\begin{remark}\label{r-1}
By using Theorem 1 in \cite{MCB} and obvious generalization of its proof for functions from the classes $L(C,T,D)$ with $T\neq 0$ one can obtain
more rapidly decreasing sequence $Y_{C,T\!,D}(r)$.
\end{remark}\smallskip

\begin{example}\label{e-1} Since the output entropy $H(\Phi(\rho))\doteq cH(\Phi(\rho)/c),\, c=\Tr\Phi(\rho)$, and the entropy exchange $H(\Phi,\rho)\doteq H(\widehat{\Phi}(\rho))$
of any  quantum operation $\Phi:A\rightarrow B$ preserving finiteness of the entropy are concave nonnegative functions on the set $\S(\H_A)$ bounded from above by the function
$H(\rho)+H^{\rm p}_{\rm max}(\Phi)$, where  $H^{\rm p}_{\rm max}(\Phi)$ is the constant defined in (\ref{HM}) \cite[Section 3.1]{PFE}, it follows from inequality (\ref{w-k-ineq}) and Lemma \ref{L-lemma} in the Appendix that both these functions belong to the class $L(1,H^{\rm p}_{\rm max}(\Phi),1)$.

Let $\mathfrak{O}_k(A,B)$ be the set of all quantum operations $\Phi$ from $A$ to $B$ such that $H^{\rm p}_{\rm max}(\Phi)\leq \log k$. The set $\mathfrak{O}_k(A,B)$ contains all  quantum operations with Choi rank not exceeding $k$ and  all quantum operations whose output has dimension not exceeding $k$. Then, by the above observation, the output entropy and the entropy exchange of any operation in
$\mathfrak{O}_k(A,B)$ belong to the class $L(1,\log k,1)$. By Theorem \ref{main} for any state $\rho$ with the FA-property we have
$$
H(\Phi(\rho_r))\mathop{\rightrightarrows}\limits_{\Phi} H(\Phi(\rho))\quad \textrm{and} \quad H(\Phi,\rho_r)\mathop{\rightrightarrows}\limits_{\Phi}  H(\Phi,\rho)
$$
as $\,r\to+\infty$, where $\mathop{\rightrightarrows}\limits_{\Phi}$ denotes the uniform convergence on the set $\mathfrak{O}_k(A,B)$ for any given $k$. By using the proof of Theorem \ref{main}
one can obtain an explicit bound for the rate of this convergence depending only on $k$ and on the spectrum of the state $\rho$.  $\square$
\end{example}\smallskip

\begin{example}\label{e-2} A quantum measurement with finite or countable outcome set $I$ is described mathematically
by a quantum instrument $\mathbf{\Upsilon}=\{\Upsilon_i\}_{i\in I}$ -- collection of quantum operations from a system $A$ to a system $A'$ such that
$\,\sum_{i\in I}\Tr\Upsilon_i(\rho)=1\,$ for any state $\rho$ in $\S(\H_A)$ \cite{H-SSQT,N&Ch,Wilde}.

The
information gain of a quantum measurement described by the instrument $\{\Upsilon_i\}_{i\in I}$ at a state $\rho$ in $\S(\H_A)$ is given by the quantum mutual information $I(R\!:\!E)$
of the state
$$
\sigma_{RE}=\sum_{i\in I}\Tr_{A'} [\shs\id_R\otimes \Upsilon_i(\hat{\rho}_{AR})]\otimes |i_E\rangle\langle i_E|,
$$
where $\hat{\rho}_{AR}$ is a purification of the state $\rho$ and $\{|i_E\rangle\}$ is a basis in a Hilbert space $\H_E$ such that $\dim\H_E=\mathrm{card}(I)$ \cite{GIB}. In  fact,
the infromation gain is completely determined by
the Positive Operator Valued Measure  (POVM) $\M_{\mathbf{\Upsilon}}=\{\Upsilon^*_i(I_{\H_{A'}})\}_{i\in I}$ corresponding to the instrument $\{\Upsilon_i\}_{i\in I}$.
Indeed, the above defined state  $\sigma_{RE}$ can be expressed as
$$
\sigma_{RE}=\Psi_{\M_{\mathbf{\Upsilon}}}\otimes \id_{R}(\hat{\rho}_{AR}),
$$
where we denote by $\Psi_{\M}$ the q-c channel $\,\rho\mapsto\sum_{i}[\Tr M_i\rho]|i_E\rangle\langle i_E|\,$ from $\S(\H_{A})$ to $\S(\H_{E})$ defined by a POVM $\M=\{M_i\}$.

So, we may consider the information gain of a quantum measurement as a function of a POVM $\M$ and a measured state $\rho$.
It is easy to see that this quantity, denoted by  $IG(\M,\rho)$ in what follows, coincides with the quantum mutual information of the channel $\Psi_{\M}$ at the state $\rho$. Hence
the function $\rho\mapsto IG(\M,\rho)$ belongs to the class $L(2,0,2)$ \cite[Section 4.3]{MCB}. By Theorem \ref{main} for any state $\rho$ with the FA-property we have
$$
IG(\M,\rho_r)\mathop{\rightrightarrows}\limits_{\M} IG(\M,\rho)
$$
as $\,r\to+\infty$, where $\mathop{\rightrightarrows}\limits_{\M}$ denotes the uniform convergence on the set of all POVM $\M$  with finite or countable outcome set. $\square$
\end{example}\smallskip

In the following section we will  show that the uniform approximation property
stated in Theorem \ref{main} holds for important characteristics of quantum channels not belonging to any of the classes $L(C,T,D)$.

\section{Uniform approximation of basic  characteristics of quantum channels}

\subsection{Characteristics depending on a pair (channel, input state)}

There are several characteristics depending on a pair (channel, input state) that play essential role in analysis of informational abilities of a quantum channel.

The \emph{constrained Holevo capacity} of a channel $\Phi:A\rightarrow B$ at a state $\rho$ is defined as
$$
\bar{C}(\Phi,\rho)=\sup_{\sum_kp_k\rho_k=\rho}\chi(\{p_k,\Phi(\rho_k)\}),
$$
where the supremum is over all ensembles $\{p_k,\rho_k\}$ of states in $\S(\H_A)$ with the average state $\rho$ \cite{H-SCI,H-Sh-1}.\footnote{$\chi(\cdot)$ denotes the Holevo quantity of an ensemble defined in (\ref{H-q}).}

The \emph{constrained one-shot private classical capacity} of a channel $\Phi:A\rightarrow B$ at a state $\rho$ is defined as
$$
\bar{C}_{\rm p}(\Phi,\rho)=\sup_{\sum_kp_k\rho_k=\rho}\chi(\{p_k,\Phi(\rho_k)\})-\chi(\{p_k,\widehat{\Phi}(\rho_k)\}),
$$
where $\widehat{\Phi}$ is any complementary channel to the channel $\Phi$ \cite{H-SCI,H-c-ch}.\smallskip

The \emph{mutual information} and the \emph{coherent information} of a quantum channel $\mathrm{\Phi}$ at a state $\rho$ in $\S(\H_A)$
are defined, respectively, as
\begin{equation*}%\label{mi-def}
 I(\mathrm{\Phi},\rho)=H(\rho)+H(\mathrm{\Phi}(\rho))-H(\mathrm{\Phi},\rho)\quad\textrm{and}\quad I_c(\mathrm{\Phi},\rho)=H(\mathrm{\Phi}(\rho))-H(\mathrm{\Phi},\rho),
\end{equation*}
where $H(\mathrm{\Phi},\rho)$ is the entropy exchange (equal to $H(\widehat{\Phi}(\rho))$) \cite{H-SCI,Wilde}.
To avoid  the uncertainty
$"\infty-\infty"$ in the infinite-dimensional case these quantities can be defined for any input state $\rho$ with finite entropy by the expressions
\begin{equation}\label{mi-def+}
 I(\mathrm{\Phi},\rho)=I(B\!:\!R)_{\mathrm{\Phi}\otimes\mathrm{Id}_{R}(\hat{\rho})}\quad \textrm{and}\quad I_c(\mathrm{\Phi},\rho)=I(B\!:\!R)_{\mathrm{\Phi}\otimes\mathrm{Id}_{R}(\hat{\rho})}-H(\rho),
\end{equation}
where $\mathcal{H}_R\cong\mathcal{H}_A$ and $\hat{\rho}\shs$
is a pure state in $\S(\H_{AR})$ such that $\hat{\rho}_{A}=\rho$ \cite{H-Sh-2}.\smallskip

According to \cite{SEoC1} the \emph{squashed entanglement} of a quantum channel $\mathrm{\Phi}$ at a state $\rho$
can be defined as
\begin{equation}\label{se-def+}
 E_{sq}(\mathrm{\Phi},\rho)=E_{sq}(\mathrm{\Phi}\otimes\mathrm{Id}_{R}(\hat{\rho})),
\end{equation}
where $\hat{\rho}$
is a pure state in $\S(\H_{AR})$ such that $\hat{\rho}_{A}=\rho$ and $E_{sq}$ in the right hand side denotes the
squashed entanglement of a state in $\S(\H_{BR})$ \cite{C&W,Tucci,SE}.
\smallskip

\begin{property}\label{qch-1} \emph{Let $\rho$ be an infinite rank state in $\S(\H_A)$ with the FA-property and
$\rho_r=[\Tr P_r\rho\shs]^{-1}P_r\rho$, where $P_r$ is the spectral projector of $\rho$ corresponding to its $r$ maximal
eigenvalues (taking the multiplicity into account). Then
\begin{equation}\label{r-conv}
f(\Phi,\rho_r) \mathop{\rightrightarrows}\limits_{\Phi}  f(\Phi,\rho),\quad f=\bar{C}, \bar{C}_{\rm p}, I, I_{\rm c}, E_{sq},
\end{equation}
as $\,r\to+\infty$, where $\,\mathop{\rightrightarrows}\limits_{\Phi}\,$ denotes the uniform convergence on the set of all channels from the system $A$ to any system $B$. In all the cases there exist bounds on the rate of convergence in (\ref{r-conv}) vanishing as $\,r\to+\infty$ that depend only on the spectrum of $\rho$.}\smallskip
\end{property}

\emph{Proof.}  Let $\rho=\sum_{i=1}^{+\infty}\lambda^{\rho}_i|\varphi_i\rangle\langle \varphi_i|$ be the spectral decomposition of the state $\rho$ and $\{g_i\}$ a non-decreasing sequence of nonnegative numbers such that (\ref{FA+}) holds. Then $P_r=\sum_{i=1}^{r}|\varphi_i\rangle\langle \varphi_i|$. Let $G$ be the positive operator on $\H_A$ defined in (\ref{DH}) satisfying condition (\ref{H-cond+}) such that
\begin{equation*}%\label{E-r}
E_{\rho}\doteq \Tr G\rho=\sum_{i=1}^{+\infty}\lambda^{\rho}_i g_i<+\infty.
\end{equation*}

Let $\{p_k,\rho_k\}$ be an arbitrary ensemble of states in $\S(\H_A)$ with the average state $\rho$.
Consider the ensemble $\{\hat{p}_k,\hat{\rho}_k\}$, where
$$
\hat{p}_k=p_k\frac{\Tr P_r\rho_k}{\Tr P_r\rho}\quad \textrm{and}\quad  \hat{\rho}_k=\frac{P_r\rho_k P_r}{\Tr P_r\rho_k}
$$
if $\Tr P_r\rho_k>0$ and it is assumed that there is no state in the $k$-th position otherwise. It is easy to see
that $\rho_r$ is the average state of the ensemble $\{\hat{p}_k,\hat{\rho}_k\}$.

By using Winter's gentle measurement lemma (cf.\cite{Wilde}) one can estimate the\break $D_0$-distance between the ensembles $\mu=\{p_k,\rho_k\}$ and  $\hat{\mu}=\{\hat{p}_k,\hat{\rho}_k\}$ (cf.\cite{CID}) as follows
$$\begin{array}{c}
2\displaystyle D_0(\mu,\hat{\mu})=\sum_k\|p_k\rho_k-\hat{p}_k\hat{\rho}_k\|_1\leq\sum_kp_k\|\rho_k-P_r\rho_k P_r\|_1+\sum_kp_k\|P_r\rho_k P_r\|_1\frac{1-\Tr P_r\rho}{\Tr P_r\rho}\\\\
\displaystyle \leq 2\sum_k p_k\sqrt{\Tr\bar{P}_r\rho_k}+\Tr\bar{P}_r\rho\leq 2\sqrt{\sum_k p_k\Tr\bar{P}_r\rho_k}+\Tr\bar{P}_r\rho\leq 3\sqrt{\Tr\bar{P}_r\rho},
\end{array}
$$
where $\bar{P}_r=I_{\H_A}-P_r$ and the third inequality follows from concavity of the function $\sqrt{x}$.

The above estimate of $D_0(\mu,\hat{\mu})$ shows that it tends to zero as $r\to+\infty$.
It is easy to see that $\Tr G\rho_r\leq E_{\rho}$ for all $r$ such that $g_r>E_{\rho}$. So, since the operator $G$
satisfies condition (\ref{H-cond+}), Proposition 7 in \cite{CID} implies that
\begin{equation}\label{chi-est-1}
|\chi(\{p_k,\Phi(\rho_k)\}-\chi\{\hat{p}_k,\Phi(\hat{\rho}_k)\}|\leq B(r)
\end{equation}
for arbitrary channel $\Phi$, where $B(r)$ is a function tending to zero as $r\to+\infty$ that is determined by the spectrum of $G$ and  by the vanishing sequence  $\Tr\bar{P}_r\rho$. It follows that
\begin{equation}\label{one}
\bar{C}(\Phi,\rho)\leq \bar{C}(\Phi,\rho_r)+B(r)
\end{equation}
for all sufficiently large $r$.

For given $r$ let $\{p_k,\rho_k\}_{k\geq1}$ be an arbitrary ensemble of states in $\S(\H_A)$ with the average state $\rho_r$.
Consider the ensemble $\{\hat{p}_k,\hat{\rho}_k\}_{k\geq 0}$, where
$$
\hat{p}_0=1-c_r, \;\;  \hat{\rho}_0=\frac{\rho-c_r\rho_r}{1-c_r}\quad \textrm{and}\quad \hat{p}_k=p_kc_r,\;\; \hat{\rho}_k=\rho_k\quad \forall k\geq 1,\quad c_r=\Tr P_r\rho
$$
($c_r\neq1$, since $\rho\neq\rho_r$). Note that $\rho$ is the average state of the ensemble $\{\hat{p}_k,\hat{\rho}_k\}_{k\geq 0}$.
For an arbitrary channel $\Phi$ we have
\begin{equation}\label{chi-est-2}
\begin{array}{c}
\displaystyle\chi(\{\hat{p}_k,\Phi(\hat{\rho}_k)\})=(1-c_r)H(\Phi(\hat{\rho}_0)\|\Phi(\rho))+c_r\sum_{k\geq 1}p_kH(\Phi(\rho_k)\|\Phi(\rho))\\\\
\displaystyle=(1-c_r)H(\Phi(\hat{\rho}_0)\|\Phi(\rho))+c_rH(\Phi(\rho_r)\|\Phi(\rho))+c_r\sum_{k\geq 1}p_kH(\Phi(\rho_k)\|\Phi(\rho_r)),
\end{array}
\end{equation}
where the second equality is obtained by using  Donald's identity \cite{Donald}. The monotonicity of the relative entropy
implies
$$
(1-c_r)H(\Phi(\hat{\rho}_0)\|\Phi(\rho))+c_rH(\Phi(\rho_r)\|\Phi(\rho))\leq(1-c_r)H(\hat{\rho}_0\|\rho)+c_rH(\rho_r\|\rho)= h_2(c_r),
$$
where $h_2$ is the binary entropy and the last equality is due to the fact that the states  $\rho_r$ and $\hat{\rho}_0$ with probabilities $c_r$ and $1-c_r$ form an ensemble of orthogonal states with the average state $\rho$. So, it follows from (\ref{chi-est-2})
and nonnegativity of the relative entropy that
\begin{equation}\label{chi-est-3}
c_r \chi(\{p_k,\Phi(\rho_k)\})\leq\chi(\{\hat{p}_k,\Phi(\hat{\rho}_k)\})\leq c_r \chi(\{p_k,\Phi(\rho_k)\})+h_2(c_r).
\end{equation}

The first inequality in (\ref{chi-est-3}) implies that
$$
\bar{C}(\Phi,\rho)\geq c_r\bar{C}(\Phi,\rho_r)\geq \bar{C}(\Phi,\rho_r)-(1-c_r)H(\rho_r),
$$
where the obvious inequality $\bar{C}(\Phi,\rho_r)\leq H(\rho_r)$ is used.

Since $H(\rho_r)\to H(\rho)<+\infty$ and $c_r\to 1$ as $r\to+\infty$, this inequality and
(\ref{one}) imply that $\bar{C}(\Phi,\rho_r)$ tends to $\bar{C}(\Phi,\rho)$ as $r\to+\infty$ uniformly on $\Phi$.\smallskip

It follows from (\ref{chi-est-3}) that
\begin{equation*}
-h_2(c_r)\leq\chi(\{p_k,\Phi(\rho_k)\})-\chi(\{\hat{p}_k,\Phi(\hat{\rho}_k)\})\leq (1-c_r)H(\rho_r).
\end{equation*}
Since these inequalities and estimate (\ref{chi-est-1}) hold for arbitrary channel $\Phi$ they imply, respectively, that
\begin{equation}\label{two}
\bar{C}_{\rm p}(\Phi,\rho_r)\leq \bar{C}_{\rm p}(\Phi,\rho)+(1-c_r)H(\rho_r)+h_2(c_r)
\end{equation}
and
\begin{equation}\label{three}
\bar{C}_{\rm p}(\Phi,\rho)\leq \bar{C}_{\rm p}(\Phi,\rho_r)+2B(r)
\end{equation}
for all sufficiently large $r$. Since $H(\rho_r)\to H(\rho)<+\infty$ and $c_r\to 1$ as $r\to+\infty$, inequalities (\ref{two}) and (\ref{three}) imply that
$\bar{C}_{\rm p}(\Phi,\rho_r)$ tends to $\bar{C}_{\rm p}(\Phi,\rho)$ as $r\to+\infty$ uniformly on $\Phi$.\smallskip

The uniform convergence of $I(\Phi,\rho_r)$ and $I_{\rm c}(\Phi,\rho_r)$ to $I(\Phi,\rho)$ and $I_{\rm c}(\Phi,\rho)$ as $r\to+\infty$
directly follows from Theorem \ref{main}, since both functions $\varrho\mapsto I(\Phi,\varrho)$ and $\varrho\mapsto I_{\rm c}(\Phi,\varrho)$ belong to the class $L(2,0,2)$
for any channel $\Phi$ \cite[Section 4.3]{MCB}.\smallskip

To show that $E_{sq}(\Phi,\rho_r)$ tends to $E_{sq}(\Phi,\rho)$ as $r\to+\infty$
uniformly on $\Phi$ consider the purification
$$
\hat{\rho}=\sum_{i,j=1}^{+\infty}\sqrt{\lambda^{\rho}_i}\sqrt{\lambda^{\rho}_j}\,|\varphi_i\rangle\langle\varphi_j|\otimes|\psi_i\rangle\langle\psi_j|
$$
of the state $\rho$ determined by a given basis $\{\psi_i\}$ in a Hilbert space $\H_R$
and the corresponding purification
$$
\hat{\rho}_r=c_r^{-1}\sum_{i,j=1}^{r}\sqrt{\lambda^{\rho}_i}\sqrt{\lambda^{\rho}_j}\,|\varphi_i\rangle\langle\varphi_j|\otimes|\psi_i\rangle\langle\psi_j|,\quad c_r=\Tr P_r\rho,
$$
of the state $\rho_r$. Let
\begin{equation}\label{H-R-def}
G_R=\sum_{i=1}^{+\infty}g_i|\psi_i\rangle\langle \psi_i|
\end{equation}
be a positive operator on $\H_{R}$, where $\{g_i\}$ is the sequence introduced at the begin of this proof. Then
\begin{equation}\label{H-R-cond++}
  \lim_{\beta \rightarrow0^+}\left[\mathrm{Tr}\, e^{-\beta G_R}\right]^{\beta}=1
\end{equation}
and
\begin{equation}\label{E-R}
\Tr G_R\hat{\rho}_R=E_{\rho}\doteq\sum_{i=1}^{+\infty}\lambda^{\rho}_i g_i<+\infty,\quad \Tr G_R[\hat{\rho}_r]_R=c_r^{-1}\sum_{i=1}^{r}\lambda^{\rho}_i g_i\leq E_{\rho}.
\end{equation}
for all $r$ such that $g_r>E_{\rho}$. By Lemma 1 in \cite{AFM}
condition (\ref{H-R-cond++}) implies that
\begin{equation}\label{F-a}
 F_{G_R}(E)\doteq\sup_{\Tr G_R\sigma\leq E}H(\sigma)=o(\sqrt{E})\quad \textrm{as} \;\; E\to+\infty
\end{equation}
where the supremum is over all states $\sigma\in\S(\H_R)$ such that $\Tr G_R\sigma\leq E$.\smallskip

Since for an arbitrary channel $\Phi$ the monotonicity of the trace norm implies
$$
\|\Phi\otimes\id_R(\hat{\rho})-\Phi\otimes\id_R(\hat{\rho}_r)\|_1\leq \varepsilon_r\doteq\|\hat{\rho}-\hat{\rho}_r\|_1,
$$
by using Proposition 22 in \cite{SE} with $\varepsilon'=\sqrt[4]{\varepsilon_r}$ and taking (\ref{E-R}) into account we obtain
$$
\begin{array}{c}
\displaystyle \left|E_{sq}(\Phi\otimes\id_R(\hat{\rho}))-E_{sq}(\Phi\otimes\id_R(\hat{\rho}_r))\right|\leq \sqrt[4]{\varepsilon_r}(1+2u_{r})F_{G_R}\!\!\left[\frac{E_{\rho}}{\sqrt[4]{\varepsilon_r}u_{r}}\right]\\\\+2g(\sqrt[4]{\varepsilon_r})+2h_2(\sqrt[4]{\varepsilon_r}u_{r}),
\end{array}
$$
for any $r$ such that $\varepsilon_r<1$, where $u_r=(1-\sqrt[4]{\varepsilon_r})/(1+\sqrt[4]{\varepsilon_r})$, $h_2$ is the binary entropy defined after (\ref{w-k-ineq}) and $g(x)=(x+1)\log(x+1)-x\log x$. Since $\varepsilon_r\to 0$ and $c_r\to 1$ as $r\to+\infty$, it follows
from (\ref{F-a}) that the r.h.s. of the above inequality tends to zero as $r\to+\infty$ (uniformly on $\Phi$).

The last assertion of the proposition follows from the above proof. $\square$

\subsection{Characteristics depending on a pair (channel, input ensemble)}

\subsubsection{The output Holevo quantity and the privacy of discrete ensembles}

Let $\mu=\{p_k,\rho_k\}$ be an arbitrary ensemble of states in $\S(\H_A)$, i.e. a finite or countable set $\{\rho_k\}$ of states in $\S(\H_A)$ with the corresponding probability distribution $\{p_k\}$, and $\Phi:\T(\H_A)\rightarrow\T(\H_B)$ be a channel from $A$ to $B$.

The output Holevo quantity $\chi_{\Phi}(\mu)$ of the channel $\Phi$ at the ensemble $\mu$  is defined as the Holevo quantity of the ensemble $\{p_k,\Phi(\rho_k)\}$, i.e.
$$
\chi_{\Phi}(\mu)\doteq\chi(\{p_k,\Phi(\rho_k)\})=\sum_{k=1}^{+\infty}p_kH(\Phi(\rho_k)\|\Phi(\bar{\rho}(\mu))),
$$
where $\bar{\rho}(\mu)=\sum_{k=1}^{+\infty}p_k\rho_k$ is the average state of $\mu$.

The privacy $\pi_{\Phi}(\mu)$ of the channel $\Phi$ at the ensemble $\mu$  is defined as
$$
\pi_{\Phi}(\mu)=\chi(\{p_k,\Phi(\rho_k)\})-\chi(\{p_k,\widehat{\Phi}(\rho_k)\}),
$$
where $\widehat{\Phi}$ is any complementary channel to the channel $\Phi$ \cite{H-SCI,H-c-ch}.

For a given countable ensemble $\mu=\{p_k,\rho_k\}_{k=1}^{+\infty}$ let $\mu_n=\{p_k/c_n,\rho_k\}_{k=1}^n$, $c_n=\sum_{k=1}^{n}p_k,$ be the ensemble of $n$ states obtained by "truncation" of $\mu$ with the corresponding renormalization of the probabilities.
It is easy to see that for a given channel $\Phi$ the quantities $\chi_{\Phi}(\mu)$
and $\pi_{\Phi}(\mu)$ are approximated by the sequences $\chi_{\Phi}(\mu_n)$
and $\pi_{\Phi}(\mu_n)$, i.e.
\begin{equation}\label{q-app}
  \chi_{\Phi}(\mu_n)\to\chi_{\Phi}(\mu)\quad \textrm{and} \quad\pi_{\Phi}(\mu_n)\to\pi_{\Phi}(\mu)\quad \textrm{as}\;\, n\to+\infty.
\end{equation}

In this subsection we show that the FA-property of the average state $\bar{\rho}(\mu)$ implies that both convergences in (\ref{q-app})
are uniform on the set of all quantum channels $\Phi$. Moreover, we show that in this case there exist bounds on the rate of convergences in (\ref{q-app}) depending only on
the total probabilities $\sum_{k>n}p_k$ of discarded states.\smallskip

\begin{property}\label{qch-2} \emph{Let $\rho$ be a state in $\S(\H_A)$ with the FA-property. There exist functions $B_{\chi}(x)$ and $B_{\pi}(x)$
vanishing as $\shs x\to 0$ such that
$$
|\chi_{\Phi}(\mu_n)-\chi_{\Phi}(\mu)|\leq B_{\chi}(\varepsilon)\quad \textrm{and}\quad  |\pi_{\Phi}(\mu_n)-\pi_{\Phi}(\mu)|\leq B_{\pi}(\varepsilon)
$$
for arbitrary channel $\Phi$ from the system $A$ to any system $B$ and arbitrary discrete ensemble $\mu=\{p_k,\rho_k\}_{k=1}^{+\infty}$ with the average state $\rho$ provided that $\sum_{k>n}p_k\leq\varepsilon$, where $\mu_n$ denotes the truncation of $\mu$ defined before. The functions $B_{\chi}(x)$ and $B_{\pi}(x)$ are determined by the spectrum of $\rho$.}
\end{property}\smallskip

\emph{Proof.} Let $\rho=\sum_{i=1}^{+\infty}\lambda^{\rho}_i|\varphi_i\rangle\langle \varphi_i|$ be the spectral decomposition of the state $\rho$ and $\{g_i\}$ a non-decreasing sequence of nonnegative numbers such that (\ref{FA+}) holds. Then the positive operator $G$ defined in (\ref{DH}) satisfies condition (\ref{H-cond+}) and
\begin{equation*}%\label{E-r}
E_{\rho}\doteq \Tr G\rho=\sum_{i=1}^{+\infty}\lambda^{\rho}_i g_i<+\infty.
\end{equation*}

The $D_0$-distance (cf.\cite{CID,MCB}) between the ensembles $\mu=\{p_k,\rho_k\}_{k=1}^{+\infty}$ and  $\mu_n=\{p_k/c_n,\rho_k\}_{k=1}^{n}$, $c_n=\sum_{k=1}^n p_k$, is easily
calculated:
$$
2D_0(\mu,\mu_0)=\sum_{k=1}^n p_k\|\rho_k/c_n-\rho_k\|_1+\sum_{k>n}\|p_k\rho_k\|_1=(c_n(1/c_n-1)+(1-c_n))=2(1-c_n).
$$
Since $c_n\bar{\rho}(\mu_n)\leq \bar{\rho}(\mu)=\rho$, we have $\Tr G\bar{\rho}(\mu_n)\leq E_{\rho}/c_n$. Hence Propositions 4 and 5 in \cite{MCB} imply
that
$$
|\chi_{\Phi}(\mu_n)-\chi_{\Phi}(\mu)|\leq \mathbb{CB}_{t}(E_{\rho}/c_n, 1-c_n\,|\,2,2)
$$
and
$$
|\pi_{\Phi}(\mu_n)-\pi_{\Phi}(\mu)|\leq \mathbb{CB}_{t}(E_{\rho}/c_n, 1-c_n\,|\,4,2)
$$
for some $t>0$ and arbitrary channel $\Phi$ from the system $A$ to any system $B$, where $\mathbb{CB}_{t}(E,\varepsilon\,|\,C,D)$ denotes the r.h.s. of continuity bound in Theorem 1
in \cite{MCB} defined via characteristics of the operator $G$. Condition (\ref{H-cond+})
implies that $\mathbb{CB}_{t}(E,\varepsilon\,|\,C,D)$ is an increasing function of $E$ tending to zero as $\varepsilon\to0$
for given $E,C,D$ and $t$ \cite{MCB}. By setting
$$
B_{\chi}(x)=\mathbb{CB}_{t}(E_{\rho}/(1-x), x\,|\,2,2)\quad \textrm{and} \quad B_{\pi}(x)=\mathbb{CB}_{t}(E_{\rho}/(1-x), x\,|\,4,2)
$$
we obtain the main assertion of the proposition.

Since the spectrum of $G$ depends only on the spectrum of $\rho$, the last assertion of the proposition follows from the above proof. $\square$

\subsubsection{The output Holevo quantity and privacy of continuous ensembles}

A \textit{generalized (continuous) ensemble} of states in $\S(\H)$ is defined as a Borel probability measure on the set $\S(\H)$ \cite{H-SCI,H-Sh-1}.
The set $\mathcal{P}(\mathcal{H})$ of all Borel probability measures on $\mathfrak{S}(\mathcal{H})$ contains the  subset $\mathcal{P}_0(\mathcal{H})$ of discrete measures corresponding to discrete ensembles. The average state of a generalized
ensemble $\mu \in \mathcal{P}(\mathcal{H})$ is defined as the barycenter of the measure
$\mu $, that is $\bar{\rho}(\mu)=\int_{\mathfrak{S}(\mathcal{H})}\rho \mu (d\rho)$.

For an ensemble $\mu \in \mathcal{P}(\mathcal{H}_{A})$ its image $\mathrm{\Phi}(\mu) $
under a quantum channel $\mathrm{\Phi}:A\rightarrow B\,$ is defined as the
ensemble in $\mathcal{P}(\mathcal{H}_{B})$ corresponding to the measure $\mu
\circ \mathrm{\Phi} ^{-1}$ on $\mathfrak{S}(\mathcal{H}_{B})$, i.e.
$\,\mathrm{\Phi} (\mu )[\mathfrak{S}_{B}]=\mu[\mathrm{\Phi} ^{-1}(\mathfrak{S}_{B})]\,$ for any Borel subset
$\mathfrak{S}_{B}$ of $\mathfrak{S}(\mathcal{H}_{B})$, where $\mathrm{\Phi} ^{-1}(\mathfrak{S}_{B})$ is the pre-image of $\mathfrak{S}_{B}$ under
the map $\mathrm{\Phi} $. If $\mu =\{p _{k},\rho _{k}\}$ then  $\mathrm{\Phi} (\mu)=\{p _{k},\mathrm{\Phi}(\rho_{k})\}$.

For a given channel $\,\mathrm{\Phi}:A\rightarrow B\,$ the output Holevo quantity of a
generalized ensemble $\mu$ in $\mathcal{P}(\H_A)$ is defined as
\begin{equation*}
\chi_{\mathrm{\Phi}}(\mu)=\int_{\mathfrak{S}(\mathcal{H})} H(\mathrm{\Phi}(\rho)\shs \|\shs \mathrm{\Phi}(\bar{\rho}(\mu)))\mu (d\rho )=H(\mathrm{\Phi}(\bar{\rho}(\mu
)))-\int_{\mathfrak{S}(\mathcal{H})} H(\mathrm{\Phi}(\rho))\mu (d\rho ),  %\label{chi-q-d+}
\end{equation*}%
where the second formula is valid under the condition $H(\mathrm{\Phi}(\bar{\rho}(\mu)))<+\infty$ \cite{H-Sh-1}.\smallskip

The privacy of a quantum channel $\mathrm{\Phi}$ at a generalized ensemble $\mu$
in $\mathcal{P}(\H_A)$ is defined as
$$
\pi_{\mathrm{\Phi}}(\mu)=\chi_{\mathrm{\Phi}}(\mu)-\chi_{\widehat{\mathrm{\Phi}}}(\mu),
$$
where $\widehat{\Phi}$ is any complementary channel to the channel $\Phi$ \cite{H-SCI,H-c-ch}.

There is a natural way to approximate a generalized ensemble $\mu$ by a sequence of
discrete ensembles with the same average state. Indeed, since the set $\mathfrak{S}(\mathcal{H})$ is separable, for each $n\in\mathbb{N}$ there exists a
sequence $\{\mathcal{A}_{k}^{n}\}_{k}$ of Borel subsets
of $\mathfrak{S}(\mathcal{H})$ with the diameter  less than $1/n$ such
that $\mathfrak{S}(\mathcal{H})=\bigcup_{k}\mathcal{A}_{k}^{n}$ and
$\mathcal{A}_{k}^{n}\cap \mathcal{A}_{l}^{n}=\emptyset $ if $l\neq
k$. We may assume w.l.o.g. that $\mu(\mathcal{A}_k)>0$ for all $k$ (otherwise, we remove the set $\mathcal{A}_k$ from the sequence).
Let
$$
p_k=\mu(\mathcal{A}_k)\quad \textrm{and} \quad \rho_k=[\mu(\mathcal{A}_k)]^{-1}\int_{\mathcal{A}_k}\varrho\mu(d \varrho).
$$
Then $\mu_n=\{p_k,\rho_k\}$ is a discrete ensemble such that $\bar{\rho}(\mu_n)=\bar{\rho}(\mu)$ for each $n$. By the arguments
from the proof of Lemma 1 in \cite{H-Sh-1} the sequence $\{\mu_n\}$ of discrete ensemble converges to the ensemble $\mu$ in the
topology of weak convergence in $\mathcal{P}(\H)$.\footnote{The weak convergence of a sequence $\{\mu_n\}\subset\mathcal{P}(\H)$ to a measure $\mu_0$ in $\mathcal{P}(\H)$ means that $\,\lim_{n\to+\infty}\int f(\rho)\mu_n(d\rho)=\int f(\rho)\mu_0(d\rho)\,$ for any continuous bounded function on $\S(\H)$ \cite{Bil,Bog,H-Sh-1}.}  \smallskip

\begin{property}\label{qch-3} \emph{Let $\mu$ be a generalized ensemble of states in $\S(\H_A)$ and $\{\mu_n\}$ the sequence of
discrete ensembles constructed by using some partition $\{\mathcal{A}_{k}^{n}\}_{k}$ of $\mathfrak{S}(\mathcal{H})$
into Borel subsets with the diameter less then $1/n$ (by the way described before the proposition). If the average state $\bar{\rho}(\mu)$  has the FA-property then there exist sequences $\{C_{\chi}(n)\}$ and $\{C_{\pi}(n)\}$
vanishing as $\,n\to +\infty$ such that
$$
|\chi_{\Phi}(\mu_n)-\chi_{\Phi}(\mu)|\leq C_{\chi}(n)\quad \textrm{and}\quad  |\pi_{\Phi}(\mu_n)-\pi_{\Phi}(\mu)|\leq C_{\pi}(n)
$$
for arbitrary channel $\Phi$ from the system $A$ to any system $B$. The sequences $C_{\chi}(x)$ and $C_{\pi}(x)$ are determined by the spectrum of $\bar{\rho}(\mu)$.}
\end{property}\smallskip

\emph{Proof.} Let $\rho=\sum_{i=1}^{+\infty}\lambda^{\rho}_i|\varphi_i\rangle\langle \varphi_i|$ be the spectral decomposition of the state $\rho=\bar{\rho}(\mu)$ and $\{g_i\}$ a non-decreasing sequence of nonnegative numbers such that (\ref{FA+}) holds. Then the positive operator $G$ defined in (\ref{DH}) satisfies condition (\ref{H-cond+}) and
\begin{equation*}%\label{E-r}
E_{\rho}\doteq \Tr G\rho=\sum_{i=1}^{+\infty}\lambda^{\rho}_i g_i<+\infty.
\end{equation*}

Denote by $D_n$ the Kantorovich distance $D_K(\mu,\mu_n)$  between the ensembles $\mu$ and $\mu_n$ \cite{O&C,CID,MCB}.
Since $\bar{\rho}(\mu_n)=\bar{\rho}(\mu)$ for each $n$, Propositions 4 and 5 in \cite{MCB} imply, respectively, that
$$
|\chi_{\Phi}(\mu_n)-\chi_{\Phi}(\mu)|\leq \mathbb{CB}_{t}(E_{\rho}, D_n\,|\,2,2)
$$
and
$$
|\pi_{\Phi}(\mu_n)-\pi_{\Phi}(\mu)|\leq \mathbb{CB}_{t}(E_{\rho}, D_n\,|\,4,2)
$$
for some $t>0$ and arbitrary channel $\Phi$ from the system $A$ to any system $B$, where $\mathbb{CB}_{t}(E,\varepsilon\,|\,C,D)$ denotes the r.h.s. of continuity bound in Theorem 1 in \cite{MCB} defined via characteristics of the operator $G$. Condition (\ref{H-cond+})
implies that $\mathbb{CB}_{t}(E,\varepsilon\,|\,C,D)$ is an increasing function of $E$ tending to zero as $\varepsilon\to 0$
for given $E,C,D$ and $t$ \cite{MCB}.

As noted above, the sequence $\{\mu_n\}$ weakly converges to the ensemble $\mu$. Hence $D_n$ tends to zero as $n\to+\infty$ \cite{Bog,CID,MCB}. Thus, by setting
$$
C_{\chi}(n)=\mathbb{CB}_{t}(E_{\rho}, D_n\,|\,2,2)\quad \textrm{and} \quad C_{\pi}(n)=\mathbb{CB}_{t}(E_{\rho}, D_n\,|\,4,2)
$$
we obtain the assertion of the proposition. $\square$

\section{On robustness of information characteristics of quantum channels}

Since quantum channels are always prepared with a finite accuracy, it is important to explore conditions under which
small perturbations of a channel $\Phi$ leads to small changes of its given information characteristic $f(\Phi,P)$, where $P$ denotes additional
parameters (quantum states, ensembles, etc.). Mathematically, this means continuity of the function $\Phi\mapsto f(\Phi,P)$ on the set of all channels between given quantum systems
equipped with some physically motivated topology (or on some its subset, for example, the subset of energy-limited channels \cite{W-EBN}).

There are significant reasons to assume that a relevant topology on the set $\F(A,B)$ of all channels between infinite-dimensional quantum systems $A$ and $B$ is the
strong convergence topology defined by the family of seminorms $\,\Phi\mapsto\|\Phi(\rho)\|$, $\rho\in\S(\H_A)$ \cite{W-EBN}. The strong convergence of a sequence $\{\Phi_n\}$
of channels to a channel $\Phi$ means that
$$
\lim_{n\to+\infty}\Phi_n(\rho)=\Phi(\rho)\quad \forall \rho\in\S(\H_A).
$$
The set $\F(A,B)$ equipped with the strong convergence topology is metrizable, and one of the metrics generating this topology on $\F(A,B)$
is the energy-constrained Bures distance $\beta_G^E$ between quantum channels induced by any positive operator $G$ on $\H_A$ with discrete spectrum of finite
multiplicity \cite{CID}.\footnote{The definition of $\beta_G^E$ is presented at the end of Section 2.}

In this section we apply general results from \cite{CID} and \cite{CSR} to obtain sufficient condition of  uniform continuity of the
basic information characteristics having the form of a function $f(\Phi, P)$, where $P$ is either an input state or an ensemble of input states, on the
set $\F(A,B)$  of all channels between arbitrary infinite-dimensional systems $A$ and $B$ w.r.t. some metric generating the strong convergence.\smallskip

\begin{property}\label{robust} \emph{Let $A$ and $B$ be arbitrary infinite-dimensional quantum systems.}
\smallskip

A) \emph{If $\rho$ is a state in $\S(\H_A)$ with the FA-property then the functions \footnote{These functions are introduced at the begin of Section 4.1.}
$$
\Phi\mapsto \bar{C}(\Phi,\rho),\quad\Phi\mapsto\bar{C}_{\rm p}(\Phi,\rho),\quad\Phi\mapsto I(\Phi,\rho),\quad \Phi\mapsto I_c(\Phi,\rho)\quad \textit{and}\quad \Phi\mapsto E_{sq}(\Phi,\rho)
$$
are uniformly continuous on the
set $\,\F(A,B)$  w.r.t. some metric generating the strong convergence
%\footnote{The uniform continuity of a function on the
%set $\,\F(A,B)$  w.r.t. the strong convergence means the uniform continuity of this function w.r.t. any metric $D$ generating the strong convergence topology on the set
%$\,\F(A,B)$ (an example of such a metric $D$ is the energy-constrained Bures distance \cite{CID}).}
and there exist uniform continuity bounds
for these functions depending only on the spectrum of $\,\rho$.}
\smallskip

B) \emph{If $\mu$ is a generalized ensemble of states in $\S(\H_A)$ with the average state $\bar{\rho}(\mu)$ having the FA-property then the functions \footnote{These functions are introduced at the begin of Section 4.2.}
$$
\Phi\mapsto \chi_{\Phi}(\mu)\quad \textit{and}\quad \Phi\mapsto \pi_{\Phi}(\mu)
$$
are uniformly continuous on the
set $\,\F(A,B)$  w.r.t. some metric generating the strong convergence. There exist uniform continuity bounds
for these functions depending only on the spectrum of $\bar{\rho}(\mu)$.}
\end{property}\smallskip

\emph{Proof.} Let $\rho=\sum_{i=1}^{+\infty}\lambda^{\rho}_i|\varphi_i\rangle\langle \varphi_i|$ be the spectral decomposition of the state $\rho$ and $\{g_i\}$ a non-decreasing sequence of nonnegative numbers such that (\ref{FA+}) holds. Then the positive operator $G$ defined in (\ref{DH}) satisfies condition (\ref{H-cond+}) and
\begin{equation*}%\label{E-r}
E_{\rho}\doteq \Tr G\rho=\sum_{i=1}^{+\infty}\lambda^{\rho}_i g_i<+\infty.
\end{equation*}
Let $E>E_{\rho}$. By Proposition 1 in \cite{CID} the energy-constrained Bures distance $\beta_G^E$ generates the strong convergence on the set $\,\F(A,B)$.

Proposition 8 in \cite{CID} implies the uniform continuity of the function $\Phi\mapsto \chi_{\Phi}(\mu)$
on the set $\,\F(A,B)$  w.r.t. the distance $\beta_G^E$ for any ensemble $\mu$ such that $\bar{\rho}(\mu)=\rho$.
This and the selective uniform continuity of the multi-valued map $\Phi\mapsto\widehat{\Phi}$ w.r.t. the distance $\beta_G^E$
proved in \cite{CSR} imply the uniform continuity of the function $\Phi\mapsto \pi_{\Phi}(\mu)\doteq \chi_{\Phi}(\mu)-\chi_{\widehat{\Phi}}(\mu)$
on the set $\,\F(A,B)$  w.r.t. the distance $\beta_G^E$ for any ensemble $\mu$ such that $\bar{\rho}(\mu)=\rho$.

The last assertion of part B of the proposition follows from the fact that the continuity bound for the function
$\Phi\mapsto \chi_{\Phi}(\mu)$ given by Proposition 8 in \cite{CID} that does not depend on $\mu$ (it depends only on the spectrum of the operator $G$ determined by the spectrum of the state $\bar{\rho}(\mu)=\rho$).

The uniform continuity of the functions $\Phi\mapsto \bar{C}(\Phi,\rho)$ and $\Phi\mapsto \bar{C}_{\rm p}(\Phi,\rho)$
on the set $\,\F(A,B)$  w.r.t. the distance $\beta_G^E$ follow from the last assertion of part B proved before.

The uniform continuity of the functions $\Phi\mapsto I(\Phi,\rho)$ and $\Phi\mapsto I_{\rm c}(\Phi,\rho)$
on the set $\,\F(A,B)$  w.r.t. the distance $\beta_G^E$ can be easily established by using Proposition 5 in \cite{CID} with $n=1$.

To show the uniform continuity of the function $\Phi\mapsto E_{sq}(\Phi,\rho)$ on the set $\,\F(A,B)$  w.r.t. the distance $\beta_G^E$
consider the purification
$$
\hat{\rho}=\sum_{i,j}\sqrt{\lambda^{\rho}_i}\sqrt{\lambda^{\rho}_j}\,|\varphi_i\rangle\langle\varphi_j|\otimes|\psi_i\rangle\langle\psi_j|
$$
of the state $\rho$ determined by a given basis $\{\psi_i\}$ in a Hilbert space $\H_R$. Let $G_R$ be
the positive operator on $\H_{R}$ defined in (\ref{H-R-def}) satisfying condition (\ref{H-R-cond++}) and $F_{G_R}$ the function defined in (\ref{F-a}).
\smallskip

Let $\Phi$ and $\Psi$ be arbitrary channels s.t. $\beta_G^E(\Phi,\Psi)\leq\varepsilon<1/2$.  Since $\Tr G\hat{\rho}_A< E$, the first inequality in (\ref{B-d-s-r}) implies
$$
\|\Phi\otimes\id_R(\hat{\rho})-\Psi\otimes\id_R(\hat{\rho})\|_1\leq 2\beta(\Phi\otimes\id_R(\hat{\rho}),\Psi\otimes\id_R(\hat{\rho}))\leq 2\beta_G^{E}(\Phi,\Psi)\leq2\varepsilon<1.
$$
Since $\Tr G_R\hat{\rho}_R<E$, by using Proposition 22 in \cite{SE} with $\varepsilon'=\sqrt[4]{2\varepsilon}$ we obtain\footnote{The assumption $\|\omega^2_{AB}-\omega^1_{AB}\|_1=\varepsilon<1$ in Proposition 22 in \cite{SE} can be replaced by the  assumption $\|\omega^2_{AB}-\omega^1_{AB}\|_1\leq\varepsilon<1$. This follows from the proof of this proposition.}
$$
\begin{array}{c}
\displaystyle \left|E_{sq}(\Phi\otimes\id_R(\hat{\rho}))-E_{sq}(\Psi\otimes\id_R(\hat{\rho}))\right|\leq \sqrt[4]{2\varepsilon}(1+2v_{\varepsilon})F_{G_R}\!\!\left[\frac{E}{\sqrt[4]{2\varepsilon}v_{\varepsilon}}\right]\\\\+2g(\sqrt[4]{2\varepsilon})+2h_2(\sqrt[4]{2\varepsilon}v_{\varepsilon}),
\end{array}
$$
where $v_{\varepsilon}=(1-\sqrt[4]{2\varepsilon})/(1+\sqrt[4]{2\varepsilon})$, $g(x)=(x+1)\log(x+1)-x\log x\,$ and $h_2$ is the binary entropy defined after (\ref{w-k-ineq}).
It follows from (\ref{F-a}) that the r.h.s. of the above inequality tends to zero as $\varepsilon\to 0$. $\square$

\section*{Appendix}

\begin{lemma}\label{L-lemma} \emph{Let $f$ be a function on $\S(\H)$ satisfying inequality (\ref{F-p-1}) then for any positive trace-non-increasing
linear map $\Phi:\T(\H)\rightarrow\T(\H)$ the function\footnote{We assume that $\,f_{\Phi}(\rho)=0\,$ if $\,\Phi(\rho)=0$.}
$$
f_{\Phi}(\rho)=\|\Phi(\rho)\|_1f\!\left(\frac{\Phi(\rho)}{\|\Phi(\rho)\|_1}\right)
$$
on $\S(\H)$ satisfies inequality (\ref{F-p-1}) with the same parameters $\,a_f$ and $\,b_f$.}
\end{lemma}\smallskip

\emph{Proof.} Let $\rho$ and $\sigma$ be arbitrary states and $\lambda$ any number in $(0,1)$.
Let $p=\|\Phi(\rho)\|_1$ and $q=\|\Phi(\sigma)\|_1$. If $pq=0$ then inequality (\ref{F-p-1}) trivially holds for the function $f_{\Phi}$. If $pq\neq 0$ then the validity of the l.h.s. of inequality (\ref{F-p-1}) for the function $f$
implies
$$
\begin{array}{rl}
\displaystyle f_{\Phi}(\lambda\rho+\bar{\lambda}\sigma)\!\!& \displaystyle =\,(\lambda p+\bar{\lambda}q)f\!\left(\frac{\lambda p}{\lambda p+\bar{\lambda}q}\frac{\Phi(\rho)}{p}
+\frac{\bar{\lambda}q}{\lambda p+\bar{\lambda}q}\frac{\Phi(\sigma)}{q}\right)\\\\
& \displaystyle \geq\, \lambda p f\!\left(\frac{\Phi(\rho)}{p}\right)+\bar{\lambda}q f\!\left(\frac{\Phi(\sigma)}{q}\right)-a_f(\lambda p+\bar{\lambda}q) h_2\!\left(\frac{\lambda p}{\lambda p+\bar{\lambda}q}\right)
\\\\
& \displaystyle =\,\lambda f_{\Phi}(\rho)+\bar{\lambda} f_{\Phi}(\sigma)-a_f (\lambda p+\bar{\lambda}q) h_2\!\left(\frac{\lambda p}{\lambda p+\bar{\lambda}q}\right),\;\, \bar{\lambda}=1-\lambda.
\end{array}
$$
So, to prove the l.h.s. of inequality (\ref{F-p-1}) for the function $f_{\Phi}$ it suffices to show
that
$$
(x+y)h_2\!\left(\frac{x}{x+y}\right)\leq  h_2(x)
$$
for all positive numbers $x$ and $y$ such that $x+y\leq 1$. This inequality follows from the concavity
of the binary entropy, since
the probability distribution $\{x, 1-x\}$ is the convex mixture of the probability distributions
$\{x/(x+y),y/(x+y)\}$ and $\{0, 1\}$ with the coefficients $\,x+y\,$ and $\,1-x-y$.

The r.h.s. of inequality (\ref{F-p-1}) for the function $f_{\Phi}$  is proved similarly. $\square$

\bigskip

I am grateful to A.S.Holevo and G.G.Amosov for the discussion that motivated this research.

\end{document}